\pdfoutput=1
\documentclass{article}
\pdfoutput=1
\usepackage[preprint,nonatbib]{neurips_2021}

\usepackage[utf8]{inputenc} % allow utf-8 input
\usepackage[T1]{fontenc}    % use 8-bit T1 fonts
\usepackage[colorlinks = true,
            linkcolor = black,
            urlcolor  = blue,
            citecolor = blue]{hyperref}       % hyperlinks
\usepackage{url}            % simple URL typesetting
\usepackage{booktabs}       % professional-quality tables
\usepackage{amsfonts}       % blackboard math symbols
\usepackage{nicefrac}       % compact symbols for 1/2, etc.
\usepackage{microtype}      % microtypography
\usepackage{graphicx} % more modern
\usepackage{amsfonts}
\usepackage{amsmath,soul}
\usepackage{amssymb}
\usepackage[capitalize]{cleveref} %cref
\usepackage{dsfont} % mathds
\usepackage[font=small,skip=0pt]{subcaption} %subfigure
\usepackage{balance}
\usepackage[font=small]{caption}
\usepackage{bm} % bold italic vectors
\usepackage{wrapfig}
\usepackage{float} % To place figure on top right as float

\usepackage{color}
\def\name/{\textit{PCE-PINNs}}

\usepackage[authormarkuptext=name,addedmarkup=bf,authormarkupposition=left]{changes}
\definechangesauthor[name={J.~H.}, color={blue}]{jh}
\definechangesauthor[name={M.~E.}, color={red}]{me}
\definechangesauthor[name={B.~L.}, color={green}]{bl}
\usepackage{tikz,mathtools}
\crefformat{equation}{(#2#1#3)}
\Crefformat{equation}{Equation~(#2#1#3)}
\Crefname{equation}{Equation}{Equations}

\usetikzlibrary{shapes,positioning,automata,arrows,fit,backgrounds,calc}
\tikzstyle{block} = [draw, fill=blue!20, rectangle,minimum height=1em,
minimum width=2em] \tikzstyle{sum} = [draw, fill=blue!20, circle, node
distance=1cm] \tikzstyle{input} = [coordinate] \tikzstyle{output} =
[coordinate] \tikzstyle{pinstyle} = [pin edge={to-,thin,black}]
\usetikzlibrary{trees} \usetikzlibrary{decorations.pathmorphing}
\usetikzlibrary{decorations.markings}
\definecolor{darkgreen}{rgb}{0,0.5,0}
\definecolor{darkred}{rgb}{220,20,60}

\newcommand{\cmmnt}[1]{\ignorespaces}

\newcommand{\bit}{\begin{itemize}}
\newcommand{\ei}{\end{itemize}}

\usepackage[bottom]{footmisc}

\title{Digital Twin Earth - Coasts: Developing a fast and physics-informed surrogate model for coastal floods via neural operators}

\author{%
Peishi Jiang$^{1}$, Nis Meinert$^{2}$, 
Helga Jordão$^{3}$, Constantin Weisser$^{4}$ \\
\textbf{Simon Holgate$^{5}$,} \textbf{Alexander Lavin,$^{6}$} \textbf{Björn Lütjens,$^{4,5}$}, \textbf{Dava Newman, $^{4}$} \\
\textbf{Haruko Wainwright$^{7}$,} \textbf{Catherine Walker$^{8,9}$,} \textbf{Patrick Barnard$^{10}$} \\
1. Pacific Northwest National Laboratory, USA \\
2. German Aerospace Center (DLR), Germany \\
3. CERENA, Universidade de Lisboa, Portugal \\
4. Massachusetts Institute of Technology, USA \\
5. IBM Research, UK \\
6. Institute for Simulation Intelligence, USA \\
7. Lawrence Berkeley National Laboratory, USA \\
8. NASA Headquarters, USA \\
9. Woods Hole Oceanographic Institution, USA \\
10. United States Geological Survey, USA \\
}

\begin{document}

\maketitle

\begin{abstract}
  Developing fast and accurate surrogates for physics-based coastal and ocean models is an urgent need due to the coastal flood risk under accelerating sea level rise, and the computational expense of deterministic numerical models.
  For this purpose, we develop the first \textit{digital twin} of Earth coastlines with new physics-informed machine learning techniques extending the state-of-art \textit{Neural Operator}.
  As a proof-of-concept study, we built Fourier Neural Operator (FNO) surrogates on the simulations of an industry-standard flood and ocean model (NEMO). 
  The resulting FNO surrogate accurately predicts the sea surface height in most regions while achieving upwards of 45x acceleration of NEMO.
  We delivered an open-source \textit{CoastalTwin} platform in an end-to-end and modular way, to enable easy extensions to other simulations and ML-based surrogate methods.
  Our results and deliverable provide a promising approach to massively accelerate coastal dynamics simulators, which can enable scientists to efficiently execute many simulations for decision-making, uncertainty quantification, and other research activities.
\end{abstract}
%keywords: time-series, neural operator, surrogate modeling, flood risk, inundation modelling, sea level rise, uncertainty quantification

%!TEX root=main.tex

\section{Introduction}
\label{sec:intro}
\vspace{-.5cm}
\begin{figure}[!th]
\centering
\includegraphics[scale=.75]{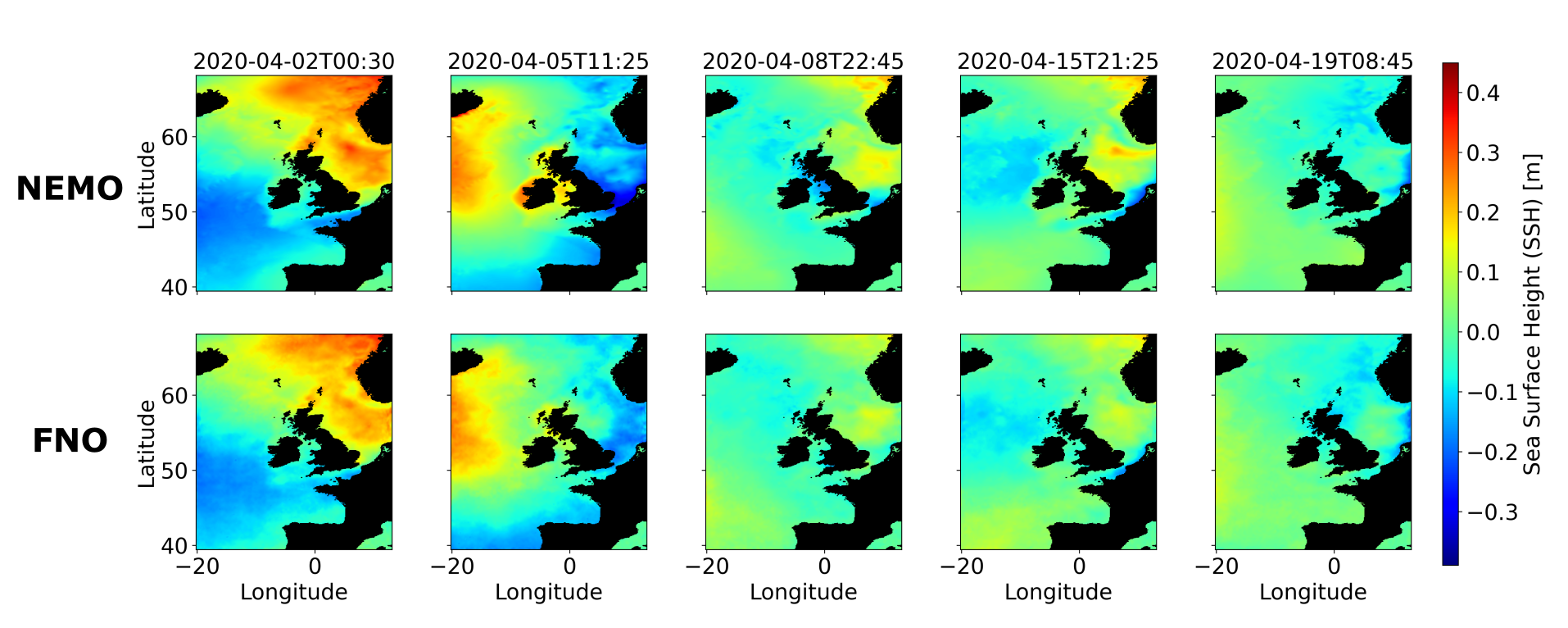}
\caption{\label{fig:1} Snapshots of NEMO simulation and FNO emulation for present time estimation using present atmospheric forcings (i.e., Case 1 in Table~\ref{tab:result}).}
\end{figure}

Coastal flooding is considered one of the most significant impacts of climate change, potentially threatening lives and damaging infrastructure with rising sea levels~\cite{Kirezci2020}. 
Increasingly into the future, coastal flooding effects on society will be exacerbated due to the increasing coastal populations and the accelerating rate of sea level rise and the severity of extreme climate events~\cite{IPCC2018, Edmonds2020}. 
Physics-based numerical models, such as Nucleus for European Modelling of the Ocean (NEMO)~\cite{NEMO}, have been developed to simulate and predict coastal and ocean dynamics.
% With a range of coastal regions and physics processes considered, 
These physical models -- driven by wind speed and mean sea level atmospheric pressure -- simulate the dynamics of water velocity and sea surface height by solving the mass and momentum conservation equations. 
Yet running these physics-based models can be extremely computationally expensive (>1 day per run), due to the need to numerically resolve multi-physics and multi-scale dynamics represented through coupled nonlinear equations in large spatial domains~\cite{Purvis2008}. 
% In particular, the simulators are not fast enough to perform simulations in an operational setting to predict coastal flooding in real-time and support emergency responses. 
In particular, these complex simulators are not fast enough for reliable risk estimation, uncertainty quantification, hypothesis testing, or real-time predictions~\cite{Purvis2008}, and are replaced by models with physical approximations that sacrifice accuracy for runtime~\cite{Bates2010}.

Machine learning (ML) methods have received much attention in the Earth Science community due to their success at providing fast data-driven models with high accuracy~\cite{Reichstein2019,Gentine2018}. 
In particular, surrogate modeling approaches replace expensive forward simulations by statistical representations through regression~\cite{Brunton2019}. 
A recent focus has been on coupling ML and physical models (such as partial differential equations (PDEs)), such that the solutions not only honor data sets but also physical constraints~\cite{Karniadakis2021, Lutjens2021}. 
However, training classical physics-informed neural networks is difficult due to the need to resolve the discretized PDE in the loss function~\cite{Raissi2019}. 
Indeed, researchers found that these approaches are unable to represent dynamics of simple cases such as a one-dimensional two-phase flow model~\cite{Fuks2020}. 
On the other hand, the recently proposed Fourier Neural Operator (FNO)~\cite{Li2021} shows a promising alternative by learning the dynamics in the frequency domain. 
In doing so, FNO is more amenable to training and directly learns the PDE-based operator which makes it mesh-independent~\cite{Li2021}.
% Among many methods, the Fourier Neural Operator (FNO) has been proposed for learning simulations from PDE-based solvers efficiently. Different from the classical physics-informed neural network~\cite{Raissi2019}, which can be computational expensive due to the need to approximate the solution of PDE , FNO is able to directly represent the PDE-based operator by learning dynamics in the frequency domain~\cite{Li2021}.
  
Here, we propose the first ``coastal digital twin'', an emulator built on state-of-art physics-informed ML techniques to produce computationally lightweight surrogate models that provide fast and accurate predictions of sea surface heights in coastal regions.  
As a proof-of-concept experiment, we developed a digital twin for the NEMO simulations in northwestern Europe using an improved version of FNO. 

Our results show: (1) the extension of FNO to learn multivariate dynamics (note that FNO was used for univariate cases in its original development~\cite{Li2021}); (2) the overall superior performance of FNO over the baseline model UNet~\cite{Ronneberger2015} in emulating sea surface height; (3) the adverse impact of masked land boundaries in training FNO; and (4) a 45x acceleration achieved by FNO compared with NEMO simulation. We deliver the code and data to reproduce these results with our open-source platform \textit{CoastalTwin}, including tools to extend our initial experiments.\footnote{Repository will be open-sourced upon publication at the following address: \href{https://gitlab.com/frontierdevelopmentlab/fdl-2021-digital-twin-coasts/coastaltwin}{gitlab.com/frontierdevelopmentlab/fdl-2021-digital-twin-coasts/coastaltwin}}
% \begin{itemize}
%     \item the capability of FNO in learning multivariate dynamics (note that FNO was used for univariate cases in its original development~\cite{Li2021});
%     \item the overall superior performance of FNO over the baseline model UNet~\cite{Ronneberger2015} in emulating sea surface height;
%     \item the adverse impact of masked land boundaries in training FNO;
%     \item a 45x acceleration achieved by FNO compared with NEMO simulation.
%     % \item the estimation divergence in longer time step forecasting.
% \end{itemize}

%!TEX root=main.tex

\section{Method}
\label{sec:method}
We first summarize the NEMO simulations and environment for this coastal climate setting, the FNO surrogate methods, and the specifications of our open-source platform \textit{CoastalTwin}. 

\begin{figure}[!h]
\centering
\includegraphics[scale=.25]{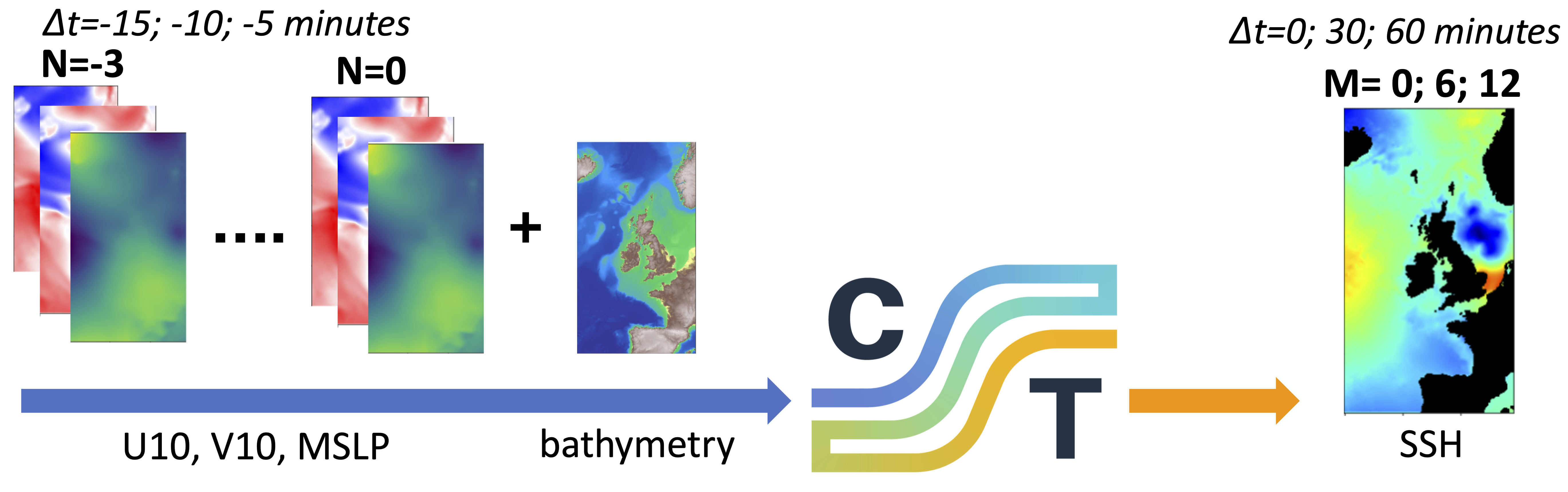}
\caption{\label{fig:2} Pipeline for emulating Sea Surface Height (SSH) based on NEMO atmospheric forcings using \textit{CoastalTwin} for the cases in Table~\ref{tab:result}.} % I love how the arrows have the same color scheme as the logo.
\end{figure}

\subsection{NEMO simulation}
\label{sec:method-1}
\vspace{-.1cm}
NEMO was set up at 7-km regular grids in northwestern Europe, composing overall 520x292 grids.
The atmospheric forcings of NEMO include mean sea level pressure (MSLP), U-direction wind speed (U10), and V-direction wind speed (V10) averaged on the top 10m above the sea surface from the downscaled product of ECMWF Reanalysis 5th Generation~\cite{Hersbach2020}. 
The bathymetry profile was from the General Bathymetric Chart of the Oceans product~\cite{GEBCO}.
The simulation of two-dimensional (2D) sea surface height (SSH) was performed for all of 2020 at every 5min.
In this experiment, we normalized the dynamic forcings and simulations (i.e., U10/V10/MSLP/SSH) to  mean zero and unit variance, and implemented a special scaling for bathymetry such that $B^\prime=\frac{\ln(B+50)-\ln(50))}{\ln(100)}$, where $B$ is the ocean depth.
% that keeps a bathymetry of zero at zero and approaches logarithmic behavior for large positive and negative values of bathymetry. 
This special scaling results in the local topological features that are sensitive to small bathymetry changes around zero, but insensitive to moderate changes at large bathymetry. 
% This preprocessing is realized by the insight that bathymetry at the coasts is more important than in the deep sea. 
We then splitted the normalized dataset into test (April 2020) and training (the remaining 11 months) datasets. 
More information on NEMO can be found at~\cite{NEMO}.

\vspace{-.15cm}
\subsection{Fourier Neural Operator}
\label{sec:method-2}
\vspace{-.2cm}
\textit{Physics-informed ML} methods integrate mathematical physics models with data-driven learning, namely with neural networks (NNs)~\cite{Karniadakis2021}. 
% More specifically, making an ML method physics-informed amounts to introducing appropriate observational, inductive, or learning biases that can steer or constrain the learning process to physically consistent solutions \cite{Karniadakis2021PhysicsinformedML}
A promising direction in spatiotemporal use-cases is \textit{neural operator learning}: using NNs to learn mesh-independent, resolution-invariant solution operators for PDEs~\cite{Lu2020, Anandkumar2020}. To achieve this, Li et al. \cite{Li2021} use a Fourier layer that implements a Fourier transform, then a linear transform, and an inverse Fourier transform for a convolution-like operation in a NN.

\vspace{-.15cm}
\subsection{CoastalTwin}
\label{sec:method-3}
\vspace{-.2cm}
For implementing FNO and other ML-based surrogates with NEMO, we developed \textit{CoastalTwin}, a modular and extensible platform to integrate simulations from physical models, such as NEMO, with ML models, to produce reliable, accelerated emulation of coastal dynamics.

Using \textit{CoastalTwin}, we developed the surrogate models of NEMO to predict SSH at time $t_M$ based on both atmospheric forcings (i.e., U10, V10, and MSLP) at preceding times $t_{N},...,t_{0}$ and the bathymetry, where $t_0$ is the present time, $N\in\mathds I$ the history and $M\in \mathds I$ the lookahead, and $\Delta t=5$min the FNO time step.
FNO is compared to a baseline UNet-based model~\cite{Ronneberger2015}.
The model was trained on various timescales constituting cases $C_i = \{N{=}0, M{=}0\}_1, \{N{=}-3, M{=}0\}_2, \{N{=}-3, M{=}6\}_3, \{N{=}-3, M{=}12\}_4$(\cref{fig:2}). While case 1 predicts the present SSH using the present forcings, cases 2-4 use forcings at purely historical time steps, $t_{N:{-1}}$, to estimate SSH at a single present or future time, $\{t_0, t_6, t_{12}\}={0,15,30}$min. Our work is the first to use FNO to represent the complexity of real-world dynamics, including multivariate, multi-scale, and coupled phenomena. Here, we simulate a coupled system of nonlinear equations including 2D momentum balance for water velocity, mass balance, and boundary conditions between ocean/sea floor/sea surface~\cite{NEMO}.
% Case-1: $M=0$ and $N=0$ for present time prediction using the current forcings. Case-2: $M=0$ and $N=-3$ for present time prediction using the historical time steps up to the past 15min. Case-3: $M=6$ and $N=-3$ for prediction at the next 30min using the historical time steps up to the past 15min. Case-4: $M=12$ and $N=-3$ for prediction at the next 60min using the historical time steps up to the past 15min.
% In each case, we trained FNO and a baseline model UNet~\cite{Ronneberger2015} using the same set of hyperparameters.

\vspace{-.15cm}
\paragraph{Modeling and experiment specs} Each FNO was developed by sequentially stacking a linear layer outputting 20 channels, 5 Fourier layers, and a final linear layer outputting 1 channel.
Each Fourier layer contains 20 channels and a maximum of 40 frequency modes in both spatial dimensions, followed by a batch normalization and ReLU activation.
Each UNet adopted three blocks of convolution in both contracting and expansive paths with the remaining architecture equivalent to~\cite{Ronneberger2015}.
We used the Adam optimization and a step-wise decreased learning scheduler with an initial rate 0.01, step size 20 epochs, and decay rate 0.1.
We trained each model using MSE as the loss function over 50 epochs and batch size 32, on one Tesla A100 Graphics Processing Unit (GPU).
We masked the land simulation in the loss to alleviate the adverse impact of land, where SSH is supposed to be zero.
In addition to MSE as a performance metric, we computed the Structural Similarity Index (SSIM)~\cite{Zhou2004} and the correlation (CORR) between times series of prediction and true at each grid point.
%!TEX root=main.tex

\section{Results}
\label{sec:results}
\vspace{-0.1cm}
\begin{table}[!ht]
\centering
\begin{tabular}{l|c|c|c|c}
\toprule
     & \multicolumn{4}{c}{MSE/1-SSIM}                   \\ \hline
     & Case 1: $N$=0,$M$=0   & Case 2: $N$=-3,$M$=0   & Case 3: $N$=-3,$M$=6   & Case 4: $N$=-3,$M$=12 \\ \hline
FNO  & 0.0011/0.2283 & 0.0018/0.2549 & 0.0011/0.2369 & 0.0011/0.2323            \\ \hline
UNet & 0.0025/0.4178 & 0.0033/0.4180 & 0.0025/0.4232 & 0.0025/0.4263             \\
\bottomrule
\end{tabular}
\caption{FNO and UNet training result of the four cases on the test dataset.}
\label{tab:result}
\vspace{-0.4cm}
\end{table}

% Comparison between FNO and UNet - General behavior
% - A table of MSE/SSIM/Averaged correlation
% Divergence behavior for longer step predictions
Table~\ref{tab:result} summarizes the experiment results. 
Our FNO approach outperforms UNet for all the four cases with respect to MSE and 1-SSIM.
This illustrates that FNO can better capture the PDE-based simulations than the baseline model, particularly in this multivariate scenario. Snapshots of FNO emulation of Case 1 on the test dataset shows its good agreement with the NEMO simulation (Fig. ~\ref{fig:1}). 
In general a significant speed-up is achieved by using the FNO surrogate, which took $\sim$2min to emulate the 1-month test dataset while the NEMO simulation took $\sim$1.5hr on a single core of a 2.6 GHz CPU -- we can expect GPU-parallelization and other optimizations to improve this speed-up another magnitude or more.
Therefore, the FNO is well posed to be used as a fast and accurate surrogate for PDE-based simulation of NEMO. 

For all the cases using FNO, the two metrics are similar to each other, with values of 0.001$\sim$0.002 and 0.228$\sim$0.255 for MSE and 1-SSIM, respectively. The close performances of Cases 1 and 2 indicates that including historical dynamics does not strictly improve the modeling performance. Meanwhile, when involving historical inputs (i.e., Cases 2-4), the results show that longer time prediction (1hr for Case 4) can be as reliable as the present prediction (Case 2). This is likely because the prediction at limited future time steps (up to 1hr) are well constrained by the PDE surrogates.

% Comparison between FNO and UNet - frequency and localization behavior
% - correlation spatial map
% - frequency coverage

\begin{figure}[!ht]
\centering
\includegraphics[scale=0.8]{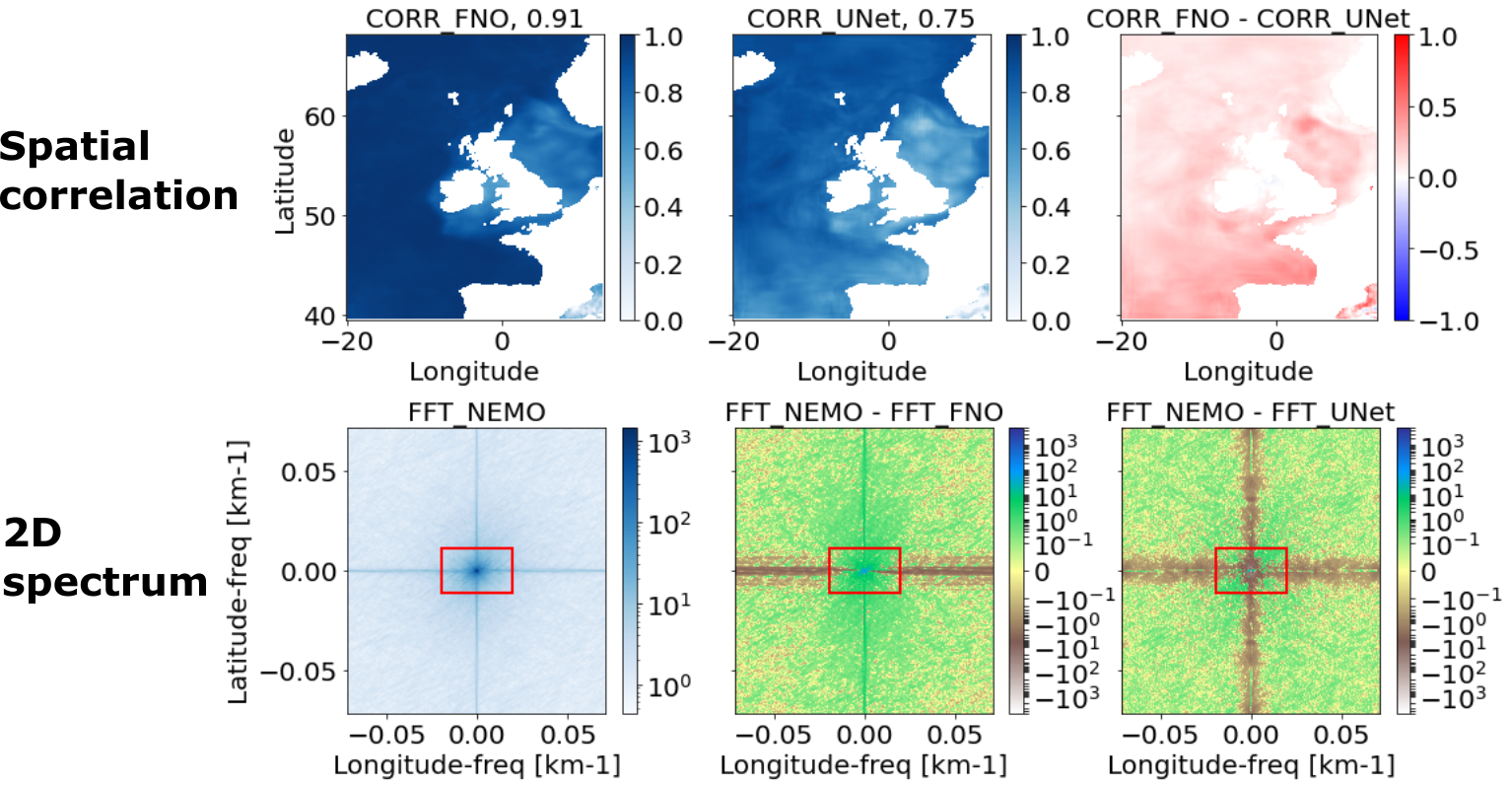}
\caption{\label{fig:3} (top) Spatial correlation between NEMO simulation and FNO/UNet emulations for Case 1 on the test dataset; (bottom) the corresponding frequency analysis using Fast Fourier Transform (FFT).}
\vspace{-.5cm}
\end{figure}

To explore the detailed estimation behavior of FNO and UNet, we plotted the spatial maps of the correlation and the 2D frequency differences between the two surrogates, based on Fast Fourier Transform (FFT). Case 1 is shown in the top panel of Fig. ~\ref{fig:3}, where we observe higher spatially averaged correlation with FNO than with UNet.
In fact, FNO predictions correlate with the NEMO simulation better in most regions than UNet except the east of France-Spain boundaries, where the SSH dynamics are severely constrained by the land surface surroundings.
The worse prediction of FNO in land-surrounded areas reveals its potential inability to resolve local dynamics that are strongly affected by masked boundaries.
Indeed, the impact of the masked boundaries is evidenced by the reduced correlations of FNO around the UK coastal region, although FNO still outperforms UNet.
The 2D frequency plot in the bottom panel of Fig. ~\ref{fig:3} shows the temporally-averaged 2D spectra of NEMO simulation, and its difference with FNO and UNet in the frequency domain.
Compared with UNet, FNO does a better job in resolving the dynamics to a great extent in the red center box where the maximum frequency cutoff of the Fourier layer is defined.
Nevertheless, both models show a significant difference with the NEMO spectra in the other cross sign frequency, which likely signifies the inability to represent coastal dynamics around the masked region.

%!TEX root=main.tex

\section{Conclusion and Outlook}
\label{sec:conclusion}
For the first time, a digital twin was developed for physics-based multivariate coastal and ocean modeling by leveraging the state-of-the-art neural operator, and demonstrated on complex real-world Earth systems data. 
Through experiments with NEMO simulations, we demonstrated the efficiency and accuracy of FNO on the sea surface height predictions, given that the training was performed with a limited dataset (i.e., single run of one-year simulation).
Future work will focus on a thorough hyperparameter tuning on FNO as well as addressing the adverse impact of masked land boundaries, which is a common issue in coastal modeling. Despite this potential limitation, it is important to note that emulation of digital twin is upwards of 45x faster than the NEMO simulation. For this reason, our coastal digital twin will be helpful for research and operation activities that require fast simulations, such as uncertainty quantification and real-time forecasting. It is nonetheless important to continue validation experiments in various settings prior to use for real-world decision making, and further investigation is suggested into downstream effects and ethical implications of such decisions.

\begin{ack}
This research was conducted at the Frontier Development
Lab (FDL), US. The authors gratefully acknowledge support from the MIT Portugal Program, National Aeronautics and Space Administration (NASA), Google Cloud, and the Coastal Observations, Mechanisms, and Predictions Across Systems and Scales (COMPASS) Project of the U.S. Department of Energy (DOE).

The authors would like to thank Zongyi Li and Anima Anandkumar from California Institute of Technology and Kamyar Azizzadenesheli from Purdue University for their insightful suggestions on the development and usage of FNO.
\end{ack}

% \section*{References}
\bibliographystyle{unsrt}
% \bibliography{references.bib}

%%%%%%%%%%%%%%%%%%%%%%%%%%%%%%%%%%%%%%%%%%%%%%%%%%%%%%%%%%%%
% \input{docs/neurips_checklist}

\appendix

% \section{Appendix}
% Optionally include extra information (complete proofs, additional experiments and plots)

\end{document}